\documentclass[conference]{IEEEtran}
\IEEEoverridecommandlockouts
\usepackage{cite}
\usepackage{amsmath,amssymb,amsfonts}
\usepackage{graphicx}
\usepackage{textcomp}
\usepackage[table]{xcolor}
\def\BibTeX{{\rm B\kern-.05em{\sc i\kern-.025em b}\kern-.08em
    T\kern-.1667em\lower.7ex\hbox{E}\kern-.125emX}}

\usepackage{stfloats}  
\usepackage{alltt}  

\usepackage{balance}

\usepackage{booktabs}  
\usepackage[bottom]{footmisc}  

\graphicspath{{fig/}{../fig/}}

\usepackage[normalem]{ulem}  

\usepackage{colortbl}

\usepackage[utf8]{inputenc}
\usepackage{paralist}  
\usepackage{xspace}  
\usepackage{url}

\definecolor{green(pigment)}{rgb}{0.0, 0.65, 0.31}

\definecolor[named]{dkred}{rgb}{0.85, 0.05, 0}

\ifdefined\RELEASE
	\newcommand{\al}[1]{} 
	
\else
	\newcommand{\al}[1]{\textcolor{green(pigment)}{[AL: #1]}} 
	
\fi

\ifdefined\ANONYMOUS
	\newcommand{\sys}{[ANONYM_BENCHMARK]\xspace}  
\else
	\newcommand{\sys}{Clubmark\xspace}  
\fi

\begin{document}

\renewcommand{\baselinestretch}{1.05}
\title{\sys: a Parallel Isolation Framework for Benchmarking and Profiling 
Clustering Algorithms
on NUMA Architectures
\thanks{
This project has received funding from the European Research Council (ERC) under the European Union's Horizon 2020 research and innovation programme (grant agreement 683253/GraphInt).
}
}


\author{
\IEEEauthorblockN{Artem Lutov, Mourad Khayati and Philippe Cudr{\'e}-Mauroux}
\IEEEauthorblockA{eXascale Infolab, University of Fribourg---Switzerland\\
Email: \{firstname.lastname\}@unifr.ch
}}

\maketitle

\begin{abstract}
There is a great diversity of clustering and community detection algorithms, which are key components of many data analysis and exploration systems. To the best of our knowledge, however, there does not exist yet any uniform benchmarking framework, which is publicly available and suitable for the parallel benchmarking of diverse clustering algorithms on a wide range of synthetic and real-world datasets. In this paper, we introduce Clubmark, a new extensible framework that aims to fill this gap by providing a parallel isolation benchmarking platform for clustering algorithms and their evaluation on NUMA servers. Clubmark allows for fine-grained control over various execution variables (timeouts, memory consumption, CPU affinity and cache policy) and supports the evaluation of a wide range of clustering algorithms including multi-level, hierarchical and overlapping clustering techniques on both weighted and unweighted input networks with built-in evaluation of several extrinsic and intrinsic measures. Our framework is open-source and provides a consistent and systematic way to execute, evaluate and profile clustering techniques considering a number of aspects that are often missing in state-of-the-art frameworks and benchmarking systems.
\end{abstract}


\section{Introduction}  
\label{sec:intro}

Clustering is a key component of many data mining systems with many applications encompassing statistical analysis and the exploration of physical, social, biological and informational systems. A wide a variety of graph algorithms have been proposed in the literature aiming to improve the efficiency and/or accuracy of the clustering. An extensive evaluation of these algorithms typically includes both real-world graphs with a ground truth as well as synthetic networks of varying parameters.
Moreover, the evaluation on synthetic networks
should 
fulfill the following desiderata:
\begin{itemize} 
\item the evaluation should be performed on various types of synthetic networks, i.e., generating networks with diverse parameters is required to check for bias in the clustering algorithm; 
\item the evaluation should take into account multiple instances of each type of synthetic network to avoid occasional bias due to particular structures in the network;
\item the evaluation should consider multiple shuffles of a given network instance to avoid bias toward a particular ordering of the input network.
\end{itemize}

The consideration of the aforementioned requirements can increase the number of input networks by orders of magnitude (i.e., by $network\_types * instances * shuffles$), which is hardly practical when using sequential execution frameworks even when considering a single clustering algorithm. In addition, parallel executions of the algorithm on multiple networks having diverse structures may affect the benchmarking results. In particular,
\begin{inparaenum}[\itshape a\upshape)]
\item the growing memory consumption of parallel processes may utilize almost all the available physical memory and cause swapping, which significantly affects execution time;
\item execution of cache-intensive processes may result in conflicting CPU cache evictions and increasing page faults, negatively affecting the execution time;
\item a bug in one algorithm or a high computational complexity on a particular network may result in interminable process executions.
\end{inparaenum}

Besides the practical considerations described above, there are also a number of important theoretical aspects that are 
missing in most existing frameworks:
%
\begin{itemize} 
\item hierarchical and multi-level algorithms may produce various numbers of output levels (resolutions), whose fair evaluation is not straightforward since an algorithm with a larger number of output levels is more likely to score high on effectiveness metrics on one of its levels;
\item very few extrinsic quality measures are applicable to overlapping clusters, which causes some frameworks to apply improper measures (e.g. ARI is used for overlaps in~\cite{Mln11}) or let end-users apply improper measures;
\item most of the quality measures for overlapping clusters are not comparable to similar measures for non-overlapping clusters (e.g. standard NMI~\cite{Dnn05} or modularity~\cite{Nwm04u} VS some overlapping NMI~\cite{Mdd11} or overlapping modularity~\cite{Zhn07,Ncs09,Lzr10} implementations), which prevents direct comparison of the respective clustering algorithms.
\end{itemize}
Finally, besides comparing to the state of the art, a benchmarking framework could be extremely useful for the online and iterative development of new clustering algorithms given interactive profiling capabilities.
To the best of our knowledge \sys is the first benchmarking framework that addresses all of the aforementioned issues and provides a unified and fully automatic solution for the comprehensive benchmarking and profiling
of diverse clustering algorithms on a wide variety of synthetic and real-world networks.

\section{Related Work}
\label{sec:relwork}
\emph{WebOCD}~\cite{Shr15} is an open-source RESTful web framework for the development, evaluation and analysis of \emph{overlapping} community detection (clustering) algorithms. It comprises several baseline algorithms, evaluation metrics and input data pre-processing utilities for the fast development of new clustering algorithms inside the framework. However, WebOCD being implemented in pure Java, is designed to execute and evaluate algorithms implemented solely in Java with specific interfaces tightly integrated into the framework. Moreover, 
the existent implementations of evaluation metrics can not be easily integrated into WebOCD without being reimplemented in Java, which is not always possible
without a significant performance drop
and time loss.

\emph{CoDAR}~\cite{Yin16} is a framework for community detection algorithm evaluation and recommendation providing a user-friendly interface and visualizations. The framework monitors the real-time structural changes of the network during the clustering process, adopts multiple metrics and builds a rating model for algorithm performance evaluation. Based on this framework, the authors also introduced a study of \emph{non-overlapping} community detection algorithms on \emph{unweighed undirected} networks~\cite{Wan15}. The evaluated algorithms are reimplemented in a common code base inside the framework, which is convenient for the uniform evaluation but limits the applicability of the framework to the existing algorithms. Unfortunately, the framework URL provided in the paper refers to a forbidden page, i.e. the implementation is not available to the public anymore.
%
%

\emph{LDBC Graphalytics}~\cite{Isp16} is a benchmark for large-scale graph analysis platforms such as Giraph and GraphX. 
It comprises several parallel algorithms, standard datasets, synthetic dataset generators, reference output and evaluation of various metrics to quantify multiple kinds of system scalability and performance variability. This benchmark provides comprehensive evaluations of graph analysis platforms on various algorithms and datasets rather than an evaluation of the algorithms themselves (i.e., evaluating the accuracy of the algorithms themselves is outside the scope of this platform).

Several frameworks and toolkits have been presented to measure the quality of the clustering, which can not be qualified as full-fledged benchmarking frameworks but are related to benchmarking.
\emph{Circulo}~\cite{Mzc15} is a framework for community detection algorithms evaluation. It executes the algorithms on preliminary uploaded input networks and then evaluates the results with multiple intrinsic and a few extrinsic measures. The framework executes the algorithms on the input datasets in parallel; however, the execution is performed without any isolation and no measures are taken to prevent mutual impact of the running processes. For example, if one of the processes consumes most of the available physical memory, others will be swapped by the operating system affecting the execution time. Multi-treaded processes in Circulo also affect the execution time of the remaining running algorithms by occupying the shared computational resources, which is unacceptable for a fair benchmarking.
A \emph{toolkit for the parallel measurement} of quality in non-overlapping clusters on both distributed and shared memory machines is presented in~\cite{Chn14}. This toolkit performs exclusively the evaluation of several intrinsic and extrinsic quality measures without executing the clustering algorithms themselves.
The evaluation of multiple algorithms using various measures is performed in several surveys and studies on clustering algorithms~\cite{Hnb14,Yan15,Hsc14}. However, the corresponding evaluation frameworks have not been publicly shared.

%


\section{System Description}
\label{sec:system}

\begin{figure*}[tbp]  
\begin{minipage}[t]{0.55\textwidth}
\centering
\includegraphics[scale=0.34]{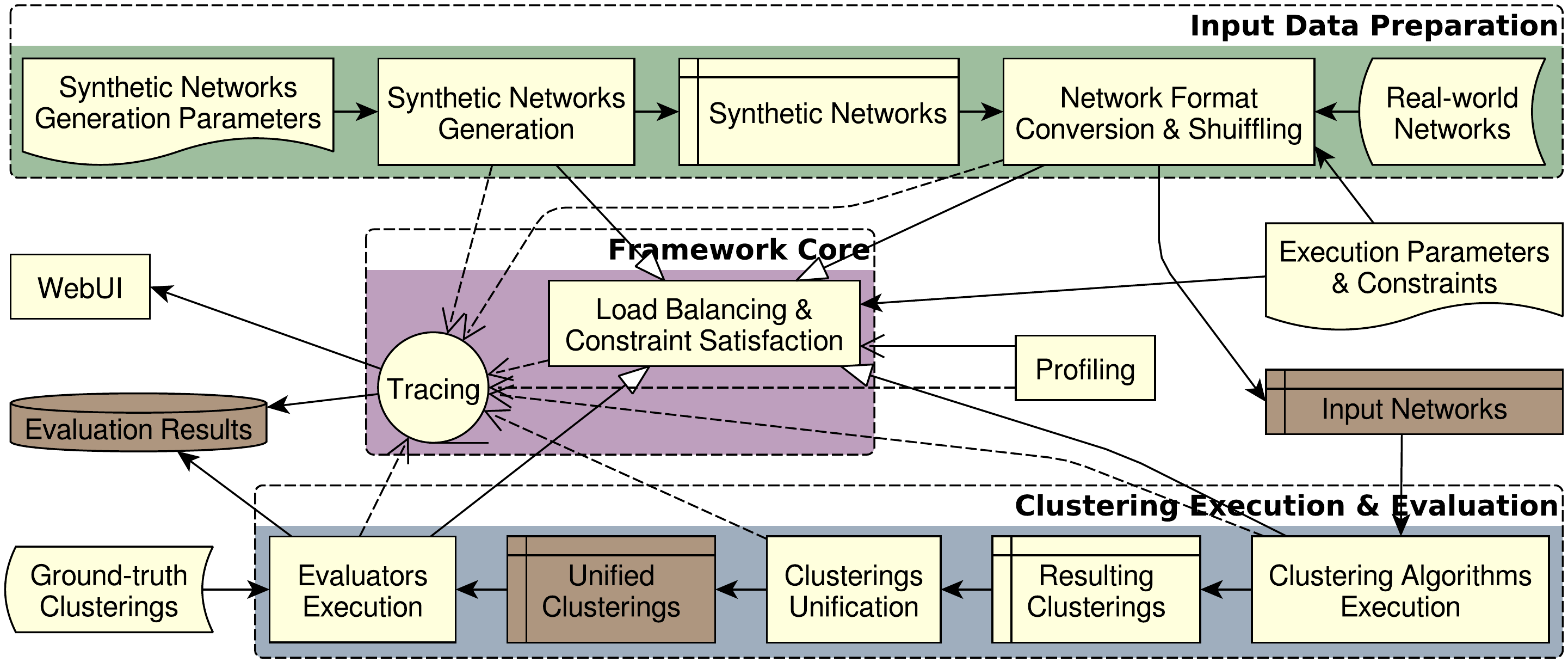}  
\caption{\sys architecture.}
\label{fig:clubmark}
%
\end{minipage}
\begin{minipage}[t]{0.45\textwidth}
\centering
\includegraphics[scale=0.315]{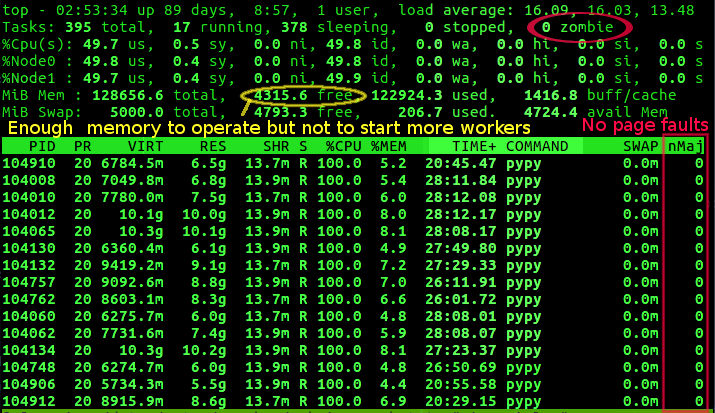}
\caption{\texttt{Top} utility listing the executing processes during benchmarking.}
\label{fig:cmdtop}
%
\end{minipage}
\end{figure*}

\sys\footnote{\url{https://github.com/eXascaleInfolab/clubmark}} is an industrial-grade benchmarking framework for 
parallel isolation benchmarking
of clustering algorithms. It can be applied on a wide variety of real-world and synthetic networks, and evaluates both the efficiency and  the effectiveness of the algorithms.
This framework is implemented in Python to be cross-platform and easily extensible, and is available as a free and open source package. \sys has a modular architecture with pluggable external clustering algorithms and utilities for the evaluation and data pre/post-processing.
Additionally, we provide a ready-to-use, containerized execution environment\footnote{\url{https://hub.docker.com/r/luaxi/clubmark-env/}} for benchmarking with pre-installed dependencies for all algorithms and utilities since most of them are implemented in compilable languages (C++ and C) with dependencies on external libraries (Boost, Intel TBB, etc.). The containerization is performed on Docker for Linux Ubuntu 16.04 LTS x64.
The overall \sys architecture is depicted in Fig.~\ref{fig:clubmark}.
%
%

\subsection{Framework Core}
\label{ssec:framecore}
The framework is based on the PyExPool\footnote{\url{https://github.com/eXascaleInfolab/PyExPool}} multi process execution pool with a constraint-aware load balancer, which isolates each executing process. PyExPool provides a number of primitives like Job, Task and ExecPool.
Each application (clustering algorithms, evaluation utilities, etc.) is scheduled for execution as a \emph{Job} instance and is transparently started under an external tiny profiler, exectime\footnote{\url{https://bitbucket.org/lumais/exectime/}}, to trace resource consumption (execution time, processing time, peak RAM consumption) and return code. A Job instance includes the execution arguments, a descriptor of the executing process, a symbolic name, an optional timeout, and provides the optional capability to restart the process on timeout. Jobs can be wrapped into a hierarchy of \emph{Tasks} for the intuitive management of related jobs. For example, a task for executing a clustering algorithm on a specific type of synthetic networks may include subtasks with several instances of this network type, where each instance may include jobs with network shuffles (randomly reordered nodes and links) of the instance.
Each Task and Job provide callbacks for the start and finish events, which can be used for the pre/post-processing activities such as notification of external services about a process crash.

All jobs are scheduled and executed by the \emph{ExecPool}. The execution pool performs load balancing to adjust the number of workers (executing processes, i.e. running jobs) in a way as to maximize the utilization of CPU and memory resources as much as possible within the specified constrains. ExecPool optionally provides isolation of the executing processes on the processing units according to the specified policy. Portable Hardware Locality utility (hwloc)\footnote{\url{https://www.open-mpi.org/projects/hwloc/}} is used to identify the hierarchical topology of the underlying NUMA system architecture to provide several policies for the maximization of the dedicated CPU L1/L2/L3 cache (process execution on the physical CPU core / node) vs parallelization (execution on a logical CPU, i.e. hardware thread). By default,
\begin{inparaenum}[\itshape a\upshape)]
\item each clustering algorithm is executed on the dedicated physical CPU core to maximize CPU L1 cache usage and to provide equal computational resources for all algorithms
\item each single-threaded evaluation utility is executed on the dedicated logical CPU to maximize parallelization
\item each instance of the multi-threaded evaluation utility (gecmi) is executed on the dedicated CPU node to maximize L1/L2/L3 cache usage and  thread parallelization.
\end{inparaenum} 
The number of worker processes in the pool is defined dynamically to satisfy
\begin{inparaenum}[\itshape a\upshape)]
\item the isolation policy and
\item considering the available hardware resources to prevent system swapping while executing the processes.
\end{inparaenum}
For example, if the isolation policy allows $n$ worker processes but $k <= n$ workers consume more than the automatically defined low memory condition\footnote{The low memory condition is defined to be a bit larger than the amount triggering a system swap, 
\texttt{vm.swappiness} is set to 5 by default.}. Then, the shortest running task among the heaviest workers (in terms of memory consumption) is killed and postponed (if $k >= 2$) as shown in Fig.~\ref{fig:cmdtop}. The execution pool also takes care of cleaning tables for the terminating processes to avoid zombies and catches, logs and handles all exceptions occurred during the execution (including  system signals), handles global timeouts and feeds the WebUI component.

\subsection{Input Data}
\label{ssec:inpdata}
\sys includes the LFR benchmark~\cite{Lnc09} 
for undirected weighted synthetic networks generation with ground-truth overlapping clusters and a script to download real-world networks with ground-truth communities from SNAP\footnote{\url{https://snap.stanford.edu/data/#communities}}.
\sys includes default parameters to generate a wide range of diverse synthetic networks with varying numbers of nodes, density of links and mixing parameters for nodes membership in overlapping clusters.
The input datasets are represented in a widely used \texttt{ncol} format. Additionally, an accessory PyNetConvert\footnote{\url{https://github.com/eXascaleInfolab/PyNetConvert}} utility is included to convert custom input networks from \texttt{pajek} and \texttt{metis} into the \texttt{nsl}\footnote{\url{https://github.com/eXascaleInfolab/clubmark/blob/master/formats}} format (generalization of \texttt{ncol}). A user may include any additional un/weighed, un/directed networks with/out ground-truth for the benchmarking from any custom location by simply specifying them in the arguments on the benchmark execution. Optionally, the specified number of shuffles is produced from each input network by reordering nodes and links to avoid bias of the  algorithms toward the input order of the data.

\subsection{Clustering Algorithms}
\label{ssec:clsalgs}
\begin{table*}[htbp]  
\centering
\caption{Clustering algorithms included in \sys.}
\label{tbl:algs}
\rowcolors{2}{gray!25}{white}
\begin{tabular}{@{}lccccccccccc@{}}
\toprule
\textbf{Features \textbackslash Algs} & \textbf{DaocA}
& \textbf{Daoc}
& \textbf{SCP} & \textbf{Louvain} & \textbf{Oslom2} & \textbf{GANXiS} & \textbf{pSCAN} & \textbf{CGGC\_RG} & \textbf{CGGCi\_RG} & \textbf{SCD} & \textbf{Randcommuns}
\\ \midrule
Hierarchical                          & +              & +             &              & +                & +               &                 &                &                   &                    &              &                      \\
Multi-scale                           & +              & +             & +            & +                & +               & +               &                &                   &                    &              &                      \\
Deterministic                         & +              & +             & +            &                  &                 &                 & +              &                   &                    &             &                      \\
Overlapping clusters                         & +              & +             & +            &                  & +               & +               & +              &                   &                    &              &                      \\
Weighted links                          & +              & +             & +            & +                & +               & +               &                &                   &                    &              & +                    \\
Parameter-free                        & +!             & +!            &              & +                & *               & *               &                & *                 & *                  & *            & +                    \\
Consensus/Ensemble                    & +              & +             &              &                  & +               &                 &                & +                 & +                  &              &                      \\ \bottomrule
\end{tabular}
\begin{flushleft}
\footnotesize{
\emph{Deterministic} means here that the algorithm is deterministic and input-order invariant;\\
+!  the feature is available and does not require any tuning, still the ability to force a manual value is provided;\\
*  the feature is parameterized and the default value is available, however a tuning might be required to obtain good results for the given network.}
\end{flushleft}
\end{table*}
Clustering algorithms can be classified by their input data type, i.e. operating on
\begin{inparaenum}[\itshape a\upshape)]
\item attributed graphs or
\item networks (graphs) specified by pairwise relations.
\end{inparaenum}
These two types of inputs cannot be unambiguously
converted into each other. Thus, the respective two types of clustering algorithms cannot be executed on the same input data and, hence, are not (directly) comparable. \sys includes a dozen of diverse clustering algorithms processing networks specified by pairwise relations. The included 
algorithms are listed in Table~\ref{tbl:algs}: SCP\footnote{\url{http://www.lce.hut.fi/research/mm/complex/software/}}~\cite{Kpl08}, Louvain\footnote{\url{http://igraph.org/c/doc/igraph-Community.html%#igraph_community_multilevel
}}~\cite{Bld08}, Oslom2\footnote{\url{http://www.oslom.org/software.htm}}~\cite{Lcn11}, 
GANXiS\footnote{\url{https://sites.google.com/site/communitydetectionslpa/}} (also known as SLPA)~\cite{Xie11}, pSCAN\footnote{\url{https://github.com/eXascaleInfolab/pSCAN}}~\cite{Chn16}, CGGC[i]\_RG\footnote{\url{https://github.com/eXascaleInfolab/CGGC}}~\cite{Ogn13} and SCD\footnote{\url{https://github.com/eXascaleInfolab/SCD}}~\cite{Prz14}.
We also included the \emph{Randcommuns}\footnote{\url{https://github.com/eXascaleInfolab/clubmark/blob/master/algorithms/%randcommuns.py
}} algorithm, which takes a number of clusters and their sizes from the ground-truth and randomly fills each formed template with connected nodes fetched from the input network. Each node is fetched only once, so some templates might end up empty and become omitted if the ground-truth clusters contain overlaps. Clusters formed by the Randcommuns represent a useful baseline for all algorithms providing interpretable and intuitive values of each evaluating measure.
%
Additional algorithms can be added just by wrapping the call of the respective application into the \texttt{exec}\textit{AlgName} Python function 
located in the \texttt{benchapps.py} module.

\subsection{Web Interface}
\label{ssec:webui}
\begin{figure*}[tbp]  
\begin{minipage}[t]{0.58\textwidth}
\centering
\includegraphics[scale=0.2]{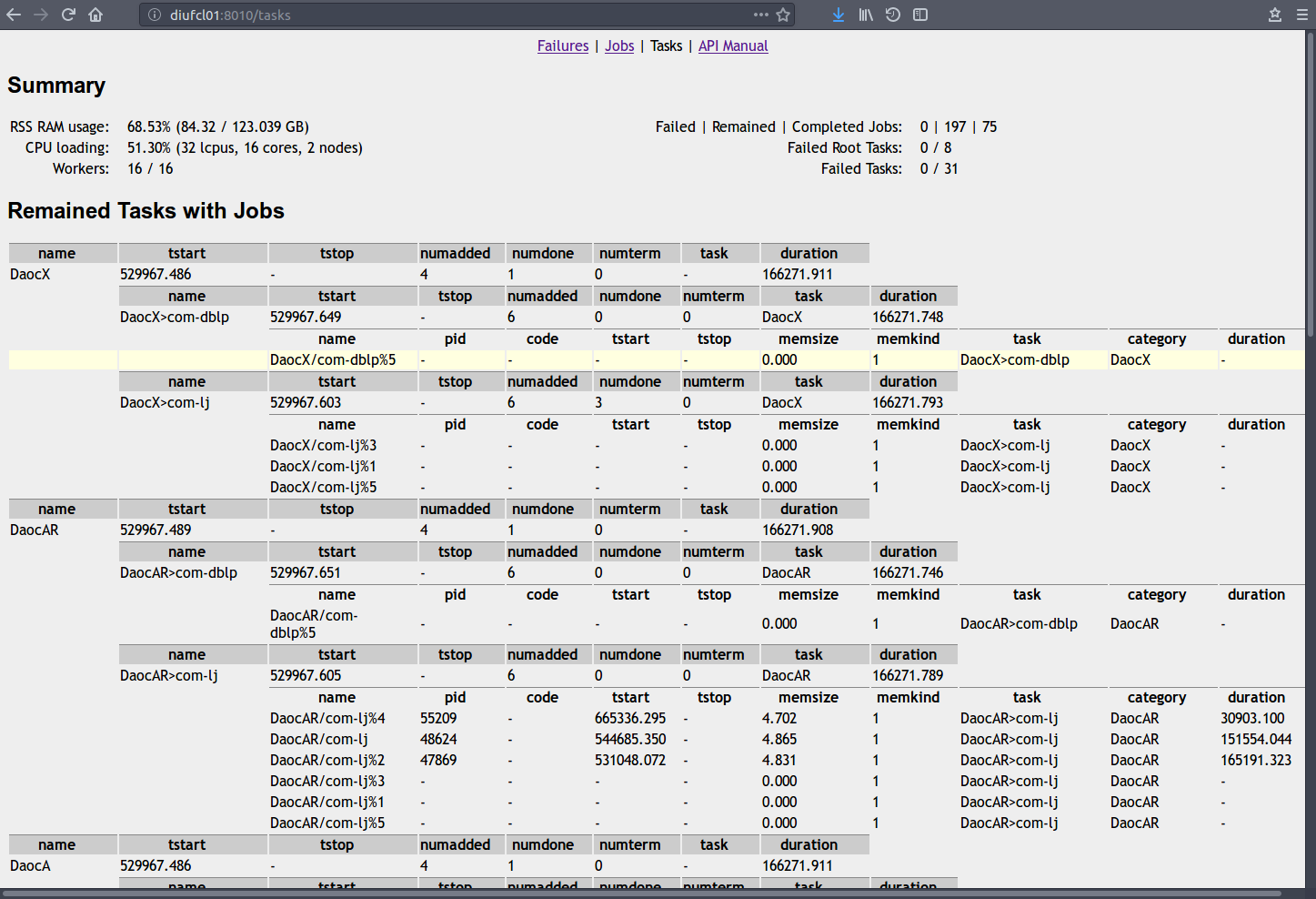}
\caption{WebUI quired from the \texttt{Firefox} browser as \texttt{http://host/tasks}.}
\label{fig:wui_tasks}
\end{minipage}
\begin{minipage}[t]{0.42\textwidth}
\centering
\includegraphics[scale=0.22]{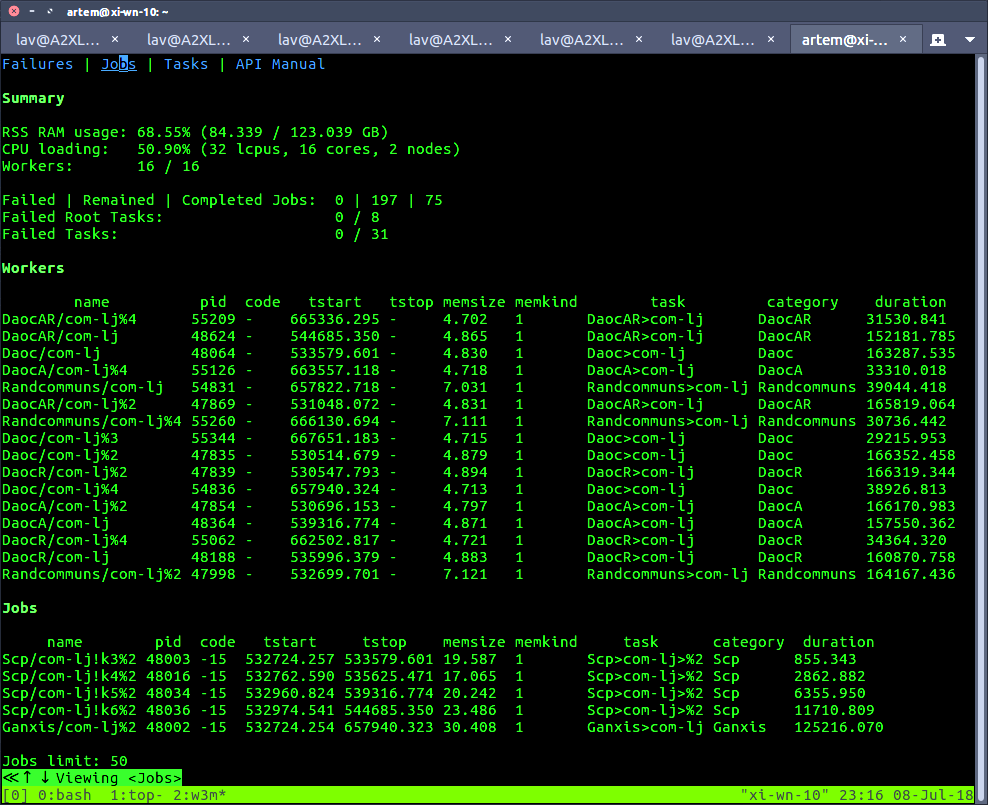}
\caption{WebUI quired from the \texttt{w3m} console browser using a simple filtering query: \texttt{\$ w3m http://host/jobs?flt=tstart}.}
\label{fig:wuic_jobs_flt}
\end{minipage}
\end{figure*}

\sys comprises a RESTful web interface (\emph{WebUI}) for the interactive profiling of the clustering algorithms and monitoring of the system resources. The WebUI provides three endpoints, each of them showing a common summary on the system resources and execution state, and also provides the following endpoint-specific information:
\begin{inparaenum}[\itshape a\upshape)]
\item \texttt{/failures} lists hierarchies of the failed tasks with their jobs
\item \texttt{/jobs} lists all executing and scheduled jobs
\item \texttt{/tasks} lists hierarchies of the executing and scheduled tasks with their jobs.
\end{inparaenum}
Each endpoint has an API for queries described at \texttt{/apinfo} and provides features such as
\begin{inparaenum}[\itshape a\upshape)]
\item queries filtering tasks and jobs using multiple fields and supporting range and optional values
\item snapshots and live continuous listing of the execution state
\item adjustment of the displayed columns for tasks and jobs 
\item selection of the outpout format (\texttt{html} or \texttt{json} for integration with other services).
\end{inparaenum}
The WebUI is built using pure HTML/CSS without any JavaScript to provide identical appearance and functionality for both the graphical and the terminal web browsers as shown in Fig.~\ref{fig:wui_tasks}-\ref{fig:wuic_jobs_flt}.

\subsection{Evaluation Measures}
\label{ssec:evalmsr}
\sys evaluates both the efficiency and the effectiveness of the clustering algorithms. The following \emph{efficiency measures} are evaluated for each algorithm:  
\begin{inparaenum}[\itshape a\upshape)]
\item peak consumption of resident memory (RAM)
\item execution time
\item processing time (total amount of time spent by the algorithm on each processing unit).
\end{inparaenum}
The \emph{effectiveness measures} comprise most common intrinsic and extrinsic clustering quality measures.
We include only the \emph{unified} measures suitable for both overlapping and non-overlapping clusters evaluation and having
a reasonably low complexity for large datasets evaluation, that is:
\begin{inparaenum}[\itshape a\upshape)]
\item having at most near linear complexity on the number of links and
\item having at most square complexity on the number of nodes.
\end{inparaenum}
Providing unified measures only prevents the improper use of measures by the end-user, facilitating a fair and direct comparison of diverse clustering algorithms.

The provided \textit{intrinsic measures} include conductance~\cite{Kan04} and modularity~\cite{Nwm04u}. As the standard modularity measure is not directly applicable to overlaps evaluation, we extend it to overlapping cases using a method for the virtual decomposition of overlaps\footnote{The description of the method is out of scope of this paper\label{ftn:outofs}, though the method is described in our open-source package}. Our decomposition technique retains the total weight and structure of the network yielding values equal to the standard modularity
when the overlap is not present.
The latter provides a fair comparison of both overlapping and non-overlapping clusters even if the non-overlapping clusters are evaluated with the standard modularity
being reported in other papers.
The provided \textit{extrinsic measures} include all known extrinsic quality measures for overlaps (fuzzy partitions)~\cite{Grg11} satisfying our complexity requirements.
In particular, Omega Index\footnote{\url{https://github.com/eXascaleInfolab/xmeasures}\label{ftn:xmeasures}}~\cite{Cln88} (which is a fuzzy version of the Adjusted Rand Index~\cite{Hbt85} and is identical to the Fuzzy Rand Index~\cite{Hlm09}), an NMI version for overlaps~\cite{Esv12} compatible with standard NMI for disjoint clustering and harmonic mean of F1-Score (F1h)\footref{ftn:xmeasures}. The average of F1-Score (F1a) is a commonly used evaluation measure of clustering accuracy~\cite{Yng13,Prat14} but the resulting values lower than $0.5$ are non-indicative since the artificial clusters formed from all permutations of the input nodes yield $F1a \rightarrow 0.5$. To make all resulting values indicative we present F1h\footref{ftn:outofs}. 
Also, we extended the original implementation of NMI for overlaps\footnote{\url{https://github.com/eXascaleInfolab/GenConvNMI}} with adaptive sampling and specific optimizations to speed up its execution, in order to apply it on large datasets.
Additional quality measures can be added by wrapping the call of the respective application into the \texttt{exec}\textit{MeasureName} Python function 
located in the \texttt{benchevals.py} module.

\subsection{Evaluation \& Results}
\label{ssec:evalres}
The uniform evaluation and fair comparison of the resulting clusterings requires some post-processing to unify their structures. In particular,
multi-level and hierarchical algorithms returing multiple levels of clusters. 
The more levels the algorithm produces, the more likely it is to have a higher evaluation. Therefore, we unify the expected number of output levels to a fixed parameter $L$ (10 by default). If the produced number of levels is larger than $L$, then the original results are moved to another repository (\texttt{-orig/}) and symbolic links are created to the $L$ output levels sampled uniformly from those results. Algorithms producing a single output level typically have a resolution parameter (density, clique rank, etc.), which can optionally be leveraged to produce $L$ output clusterings.

Thus, each clustering algorithm produces up to $L$ clusterings for each shuffle of each network instance of each network type forming the following structure of output directories:
\begin{alltt}\small
<algname>/
  <nettype>[^<instance>]/
    <shuffle>/
\end{alltt}
\noindent with up to $L$ clusterings in the \texttt{<shuffle>/}. The \textit{ground-truth} is specified per network instance. The quality evaluation of the resulting clusterings is performed by the measures specified by the user and saved into an HDF5 hierarchical structure. Afterwards, the aggregated final results are computed for each measure: 
\begin{inparaenum}[\itshape 1\upshape)]
\item the average value and variance are calculated for all shuffles of each instance and
\item these results are averaged for all instances of each network type.
\end{inparaenum}
The efficiency evaluations reported by the \texttt{exectime} for each clustering algorithm execution on each shuffle are aggregated similarly to the quality evaluations.

\section{Demonstration Objectives}
\label{sec:demo}

\sys benchmarking framework is open sources and is available for free for both non-commercial and commercial purposes from \url{https://github.com/eXascaleInfolab/clubmark}.
The objective of our demonstration is to let the audience experience the deployment, execution and extension of our proposed benchmarking framework, as well as the online profiling of a clustering algorithms via the provided web interface.
The resulting experience should let the user understand how \sys helps in
\begin{inparaenum}[\itshape a\upshape)]
\item facilitating comprehensive evaluations of emerging clustering algorithms, and
\item speeding up their development thanks to fast and convenient profiling tools on diverse input networks.
\end{inparaenum}

\textbf{\sys Deployment.}
We will demonstrate the deployment of \sys on cloud-hosted servers in two scenarios:
\begin{inparaenum}[\itshape a\upshape)]
\item directly on the host and
\item using a containerized environment from the Docker hub.
\end{inparaenum}

\textbf{Benchmarking Execution.}
We will demonstrate a typical execution of the benchmarking cycle for the clustering algorithms selected by the user on both synthetic and a predifined set of real-world networks. The synthetic networks will be generated using default framework configurations. We plan to discuss and demonstrate all essential aspects of our framework. The participants will hence experience:
\begin{inparaenum}[\itshape a\upshape)]
\item how to apply optional pre-processing steps (i.e. synthetic networks generation and shuffling)
\item how to specify execution constraints for the benchmarking
\item how a uniform input is created from distinct datasets, considering the availability of distinct instances and shuffles for each network
\item what kind of clusterings are generated by the diverse clustering algorithms
\item how results of the diverse clustering algorithms are unified for the subsequent evaluation
\item how the parallel evaluation is performed considering both single and multi-threaded applications
\item how the evaluation results are aggregated and finally stored.
\end{inparaenum}

We will also demonstrate several built-in utilities to illustrate
\begin{inparaenum}[\itshape a\upshape)]
\item the load of the server during execution
\item the structure of the input and intermediate results
\item the structure of the generating logs
\item the structure of the efficiency results produced by an external application (\texttt{exectime})
\item the structure of the final results in the HDF5 storage.
\end{inparaenum}

\textbf{Algorithms Profiling.}
We will demonstrate the available endpoints of the Web interface and encourage users to perform various queries to
\begin{inparaenum}[\itshape a\upshape)]
\item experience the REST WebAPI and its capabilities
\item perform a simple profiling of the algorithms being executed
\item understand how load-balancing and isolation work for clustering algorithms executed in parallel
\item monitor for potential issues and failed tasks when executing clustering algorithms.
\end{inparaenum}

\textbf{Framework Extensions.}
Finally, we plan to demonstrate how to seamlessly extend \sys with
\begin{inparaenum}[\itshape a\upshape)]
\item new clustering algorithms,
\item multiple versions of a single clustering algorithm (which is a useful feature when designing a new clustering algorithm) and
\item new evaluation measures
\end{inparaenum}
implemented as external applications and in various languages.

\medskip
This overall demonstration will make users aware of the latest methodologies and common pitfalls in clustering evaluation, and present an extensive overview of our \sys platform for performance evaluation and profiling.

\bibliographystyle{IEEEtran}\balance
\bibliography{IEEEabrv,clubmark}

\begin{thebibliography}{10}
\providecommand{\url}[1]{#1}
\csname url@samestyle\endcsname
\providecommand{\newblock}{\relax}
\providecommand{\bibinfo}[2]{#2}
\providecommand{\BIBentrySTDinterwordspacing}{\spaceskip=0pt\relax}
\providecommand{\BIBentryALTinterwordstretchfactor}{4}
\providecommand{\BIBentryALTinterwordspacing}{\spaceskip=\fontdimen2\font plus
\BIBentryALTinterwordstretchfactor\fontdimen3\font minus
  \fontdimen4\font\relax}
\providecommand{\BIBforeignlanguage}[2]{{%
\expandafter\ifx\csname l@#1\endcsname\relax
\typeout{** WARNING: IEEEtran.bst: No hyphenation pattern has been}%
\typeout{** loaded for the language `#1'. Using the pattern for}%
\typeout{** the default language instead.}%
\else
\language=\csname l@#1\endcsname
\fi
#2}}
\providecommand{\BIBdecl}{\relax}
\BIBdecl

\bibitem{Mln11}
V.~Melnykov and R.~Maitra, ``Carp: Software for fishing out good clustering
  algorithms,'' \emph{J. Mach. Learn. Res.}, vol.~12, pp. 69--73.

\bibitem{Dnn05}
L.~Danon, A.~D{\'{\i}}az-Guilera, J.~Duch, and A.~Arenas, ``{Comparing
  community structure identification},'' \emph{Journal of Statistical
  Mechanics: Theory and Experiment}, vol.~9, p.~8, Sep. 2005.

\bibitem{Nwm04u}
M.~E.~J. Newman and M.~Girvan, ``Finding and evaluating community structure in
  networks,'' \emph{Phys. Rev. E}, vol.~69, no.~2, p. 026113, 2004.

\bibitem{Mdd11}
A.~F. McDaid, D.~Greene, and N.~J. Hurley, ``Normalized mutual information to
  evaluate overlapping community finding algorithms,'' \emph{CoRR}, vol.
  abs/1110.2515, 2011.

\bibitem{Zhn07}
S.~Zhang, R.-S. Wang, and X.-S. Zhang, ``{Identification of overlapping
  community structure in complex networks using fuzzy c-means clustering},''
  \emph{Physica A: Statistical Mechanics and its Applications}, vol. 374,
  no.~1, pp. 483--490, 2007.

\bibitem{Ncs09}
V.~Nicosia, G.~Mangioni, V.~Carchiolo, and M.~Malgeri, ``Extending the
  definition of modularity to directed graphs with overlapping communities,''
  \emph{J Stat Mech.}, vol.~3, p.~24, Mar. 2009.

\bibitem{Lzr10}
A.~L{\'a}z{\'a}r, D.~Abel, and T.~Vicsek, ``Modularity measure of networks with
  overlapping communities,'' \emph{EPL (Europhysics Letters)}, vol.~90, no.~1,
  p. 18001, 2010.

\bibitem{Shr15}
M.~Shahriari, S.~Krott, and R.~Klamma, ``Webocd: A restful web-based
  overlapping community detection framework,'' ser. i-KNOW '15.\hskip 1em plus
  0.5em minus 0.4em\relax ACM.

\bibitem{Yin16}
X.~Ying, C.~Wang, M.~Wang, J.~X. Yu, and J.~Zhang, ``Codar: Revealing the
  generalized procedure \& recommending algorithms of community detection,''
  ser. SIGMOD '16.\hskip 1em plus 0.5em minus 0.4em\relax ACM, 2016, pp.
  2181--2184.

\bibitem{Wan15}
M.~Wang, C.~Wang, J.~X. Yu, and J.~Zhang, ``Community detection in social
  networks: An in-depth benchmarking study with a procedure-oriented
  framework,'' \emph{Proc. VLDB Endow.}, vol.~8, pp. 998--1009, Jun. 2015.

\bibitem{Isp16}
A.~Iosup, T.~Hegeman, W.~L. Ngai, S.~Heldens, A.~Prat-P{\'e}rez, T.~Manhardto,
  H.~Chafio, M.~Capot\u{a}, N.~Sundaram, M.~Anderson, I.~G. T\u{a}nase, Y.~Xia,
  L.~Nai, and P.~Boncz, ``Ldbc graphalytics: A benchmark for large-scale graph
  analysis on parallel and distributed platforms,'' \emph{Proc. VLDB Endow.},
  vol.~9, pp. 1317--1328, Sep. 2016.

\bibitem{Mzc15}
\BIBentryALTinterwordspacing
P.~Mazzuca and Y.~T, ``{Circulo: A Community Detection Evaluation Framework},''
  Feb. 2015. [Online]. Available:
  \url{https://www.lab41.org/circulo-a-community-detection-evaluation-framework/}
\BIBentrySTDinterwordspacing

\bibitem{Chn14}
M.~Chen, S.~Liu, and B.~K. Szymanski, ``Parallel toolkit for measuring the
  quality of network community structure.'' in \emph{ENIC}, 2014, pp. 22--29.

\bibitem{Hnb14}
S.~Harenberg, G.~Bello, L.~Gjeltema, S.~Ranshous, J.~Harlalka, R.~Seay,
  K.~Padmanabhan, and N.~Samatova, ``Community detection in large-scale
  networks: A survey and empirical evaluation,'' \emph{WIREs Comput. Stat.},
  vol.~6, pp. 426--439, Nov. 2014.

\bibitem{Yan15}
J.~Yang and J.~Leskovec, ``Defining and evaluating network communities based on
  ground-truth,'' \emph{Knowl. Inf. Syst.}, vol.~42, pp. 181--213, Jan. 2015.

\bibitem{Hsc14}
M.~Hasco{\"e}t and G.~Artignan, ``{Evaluation of Clustering Algorithms: a
  methodology and a case study},'' Tech. Rep. RR-14008, Sep. 2014.

\bibitem{Lnc09}
A.~Lancichinetti and S.~Fortunato, ``Benchmarks for testing community detection
  algorithms on directed and weighted graphs with overlapping communities.''
  \emph{Phys. Rev. E}, vol. 80 1 Pt 2, p. 016118, 2009.

\bibitem{Kpl08}
J.~M. Kumpula, M.~Kivel{\"a}, K.~Kaski, and J.~Saram{\"a}ki, ``Sequential
  algorithm for fast clique percolation,'' \emph{Phys. Rev. E}, vol.~78, no.~2,
  p. 026109, 2008.

\bibitem{Bld08}
V.~D. Blondel, J.-L. Guillaume, R.~Lambiotte, and E.~Lefebvre, ``Fast unfolding
  of communities in large networks,'' \emph{J Stat Mech.}, vol. 2008, no.~10,
  p. P10008, oct 2008.

\bibitem{Lcn11}
A.~{Lancichinetti}, F.~{Radicchi}, J.~J. {Ramasco}, and S.~{Fortunato},
  ``{Finding Statistically Significant Communities in Networks},'' \emph{PLoS
  ONE}, vol.~6, p. 18961, Apr. 2011.

\bibitem{Xie11}
J.~Xie, B.~K. Szymanski, and X.~Liu, ``Slpa: Uncovering overlapping communities
  in social networks via a speaker-listener interaction dynamic process,'' in
  \emph{ICDM 2011 Workshop on DMCCI}, pp. 344--349.

\bibitem{Chn16}
L.~Chang, W.~Li, X.~Lin, L.~Qin, and W.~Zhang, ``pscan: Fast and exact
  structural graph clustering,'' in \emph{ICDE'16}, pp. 253--264.

\bibitem{Ogn13}
M.~Ovelgönne and A.~Geyer-Schulz, ``An ensemble learning strategy for graph
  clustering,'' in \emph{Graph Partitioning and Graph Clustering - 10th DIMACS
  Workshop}, vol. 588, 2013, pp. 187--206.

\bibitem{Prz14}
A.~Prat-P{\'e}rez, D.~Dominguez-Sal, and J.-L. Larriba-Pey, ``High quality,
  scalable and parallel community detection for large real graphs,'' ser. WWW
  '14.\hskip 1em plus 0.5em minus 0.4em\relax ACM, 2014, pp. 225--236.

\bibitem{Kan04}
R.~Kannan, S.~Vempala, and A.~Vetta, ``On clusterings: Good, bad and
  spectral,'' \emph{J. ACM}, vol.~51, no.~3, pp. 497--515, May 2004.

\bibitem{Grg11}
S.~Gregory, ``Fuzzy overlapping communities in networks,'' \emph{J Stat Mech.},
  vol. 2011, no.~02, p. P02017, 2011.

\bibitem{Cln88}
L.~M. Collins and C.~W. Dent, ``{Omega: A general formulation of the rand index
  of cluster recovery suitable for non-disjoint solutions},''
  \emph{Multivariate Behavioral Research}, vol.~23, no.~2, pp. 231--242, 1988.

\bibitem{Hbt85}
L.~Hubert and P.~Arabie, ``Comparing partitions,'' \emph{Journal of
  Classification}, vol.~2, no.~1, pp. 193--218, Dec 1985.

\bibitem{Hlm09}
E.~Hullermeier and M.~Rifqi, ``A fuzzy variant of the rand index for comparing
  clustering structures,'' ser. IFSA-EUSFLAT 2009.

\bibitem{Esv12}
A.~V. Esquivel and M.~Rosvall, ``Comparing network covers using mutual
  information,'' \emph{CoRR}, vol. abs/1202.0425, 2012.

\bibitem{Yng13}
J.~Yang and J.~Leskovec, ``Overlapping community detection at scale: A
  nonnegative matrix factorization approach,'' ser. WSDM '13.\hskip 1em plus
  0.5em minus 0.4em\relax ACM, pp. 587--596.

\bibitem{Prat14}
A.~Prat-P{\'e}rez, D.~Dominguez-Sal, and J.-L. Larriba-Pey, ``High quality,
  scalable and parallel community detection for large real graphs,'' ser.
  WWW'14.\hskip 1em plus 0.5em minus 0.4em\relax ACM, pp. 225--236.

\end{thebibliography}

\end{document}